# Enhanced Emittance Evaluation using 2D Transverse Phase Space Distributions, High Resolution Image Denoising, and Deep Learning

F.R. Osswald[1]*, M. Chahbaoui[2] and X. Liang[3]

[1]IPHC, CNRS Nucléaire et Particules, Université de Strasbourg, 67000, Strasbourg, France.
[2]Université de Strasbourg, 67000, Strasbourg, France.
[3]Sorbonne Université, 75006 Paris, France.

* Corresponding author. E-mail: francis.osswald@iphc.cnrs.fr

***Abstract***

Next‑ generation particle accelerators demand advanced beam‑ diagnostic capabilities to ensure high performance, operational reliability, and sustainable machine operation. Increasing beam intensities and stored energies make the precise characterization of transverse profiles, phase‑ space distributions, and halos often five orders of magnitude below the core but with significant environmental impact essential for loss mitigation and machine protection. Traditional analysis methods struggle with the heterogeneous, noisy, and non‑ Gaussian data produced under realistic operating conditions. This work presents a novel tool based on an unsupervised deep‑ convolutional neural‑ network framework that significantly enhances image denoising and restoration for emittance measurements. The method reconstructs beam‑ halo structures with unprecedented resolution, detecting signals at radii beyond seven standard deviations and particle densities below $10^{-4}$ of the total beam intensity. Despite very low signal‑ to‑ noise ratios and small, non‑ annotated datasets, the approach preserves fine structures and reveals halo features previously unobserved at pilot installations. Built on a U‑ Net architecture with tailored early‑ stopping strategies and physics‑ informed metrics, the framework operates entirely on CPUs and requires minimal computational resources. The results demonstrate the potential of unsupervised deep learning as an enabling technology for high‑ dynamic‑ range beam diagnostics and motivate further development of systematic benchmarking and physics‑ informed learning strategies.

# 1 Introduction

In high‑ energy particle colliders, large‑ scale research infrastructures central to both fundamental and applied physics, the development of advanced beam diagnostics has become a decisive factor for achieving the performance, reliability, and sustainability required by next‑ generation accelerators. As beam intensities, stored energies, and facility dimensions continue to increase, the precise control and minimization of beam losses are now essential not only for machine protection and operational efficiency but also for reducing activation, limiting environmental impact, and enabling long‑ term sustainable operation. Meeting these objectives demands highly accurate, high‑ dynamic‑ range measurements of key beam parameters, including transverse position, spatial profiles, and phase‑ space distributions. In particular, the characterization of extremely low‑ intensity beam halos often five orders of magnitude below the core signal poses significant challenges in terms of charge detection sensitivity, background suppression, and robustness of data interpretation. The difficulty of validating diagnostic performance under realistic operating conditions further reinforces the need for innovative approaches. This evolution mirrors trends observed in other technologically demanding fields such as medical imaging, satellite‑ based Earth observation, autonomous‑ vehicle driving, and identification of particles in high-energy physics, where increasingly complex systems rely on sophisticated data processing pipelines to extract reliable information from noisy or incomplete measurements. Traditional analysis methods are no longer sufficient to meet the precision, speed, and automation requirements of modern and future accelerator facilities [1-3]. In this context, advanced image‑ processing techniques and artificial‑ intelligence‑ based methods offer promising avenues for enhancing data cleaning, corrupted image restoration, and denoising. In this paper, the discussion focuses on the physical role and operational relevance of beam distributions and halo formation, with particular emphasis on novel data‑ driven noise‑ reduction and image‑ based analysis techniques essential for extracting reliable information from increasingly complex and high‑ dynamic‑ range measurements. As accelerator facilities push toward higher intensities, tighter tolerances, and more demanding operational regimes, the ability to reconstruct accurate transverse profiles and halo populations becomes a prerequisite for machine protection, loss mitigation, and luminosity optimization.

Our review highlights the limitations of conventional statistical approaches when confronted with non‑Gaussian distributions, irregular phase‑space structures, and fluctuating background conditions, and examines how modern image‑processing pipelines based on convolutional architectures and denoising models can address these shortcomings. Beyond summarizing existing techniques, the work also outlines the broader context in which these developments are taking place. As in other data‑intensive fields, accelerator diagnostics are undergoing a transition toward more adaptive, robust, and automated analysis frameworks. This evolution is driven by the need to handle heterogeneous datasets, variable noise sources, and complex beam dynamics that cannot be adequately captured by traditional models.

## 2 Physics context

### 2.1 Beam distributions

Beam‑quality diagnostics such as transverse profile, beam position, and emittance measurements are essential to the control, optimization, and safe operation of particle accelerators. These quantities directly determine key performance metrics including luminosity, brightness, and transmission efficiency. A particularly critical aspect of beam characterization is the description of the transverse halo, which arises from the population of particles located in the tails of the transverse distribution, see Fig. 1. Halo formation is driven by a combination of optics mismatch, nonlinear field components, beam dynamics, collective effects, and complex beam–environment interactions. The boundary between the beam core and halo is commonly associated with the location of the maximum variation in the profile slope corresponding to the peak of the second derivative in a given transverse plane. Although halo densities are often approximated by Gaussian‑like tails, experimental observations reveal a wide variety of shapes and structures, reflecting the diversity of underlying physical mechanisms. They extend well beyond the core of the beam [4-8], but due to the number and complexity of interactions with the environment, they can take on very different appearances [9-12]. Despite representing a small fraction of the total beam population, typically less than 10%, halo particles may occupy more than 70% of the transverse phase‑space area. This disproportionate spatial footprint makes the halo the dominant contributor to beam–aperture interactions, uncontrolled losses, and component activation. Emittance growth driven by tail enhancement further exacerbates these effects and necessitates accurate six‑dimensional beam matching across successive accelerator sections [13]. Proper matching reduces the occupied phase‑space volume, increases beam brightness, and mitigates loss mechanisms that can otherwise trigger cascading failures such as overheating of equipment, material sputtering, superconducting‑cavity quenching, melting, or excessive activation.

The operational implications are particularly severe for high‑power and high‑energy machines. For example, the future HL‑LHC will store 680 MJ of beam energy, with the halo expected to contain around 35 MJ corresponding to a few percent of the total beam current located beyond three standard deviations of a Gaussian core. Recent operational feedback indicates that beam scraping will need to be performed at eight standard deviations (compared to five previously) to account for transverse and longitudinal fluctuations, bunch‑to‑bunch variations, and turn‑by‑turn evolution [14-15]. Similarly, the CADS project at J‑PARC aims to deliver 30 MW of proton power, requiring scrapers positioned at fifteen standard deviations to mitigate the impact of a 1% halo fraction on quadrupole magnets. Under such conditions, static and dynamic beam‑position errors arising from alignment tolerances or steering imperfections must be controlled to maintain losses below 1 W/m [16].

Accurate measurement of halo distributions requires detecting electric charges over a dynamic range spanning up to six orders of magnitude [17-20]. One of the fundamental challenges is the separation of true halo signal from noise in the absence of a reliable prior model of the noise distribution. This difficulty is compounded by fluctuating experimental conditions, variable background levels, and the presence of parasitic beams, all of which hinder the development of universal denoising strategies. Classical statistical approaches remain widely used in accelerator facilities, but they exhibit significant limitations when confronted with irregular beam shapes, non‑Gaussian or non‑elliptical phase‑space distributions, non‑uniform densities, or complex contamination patterns.

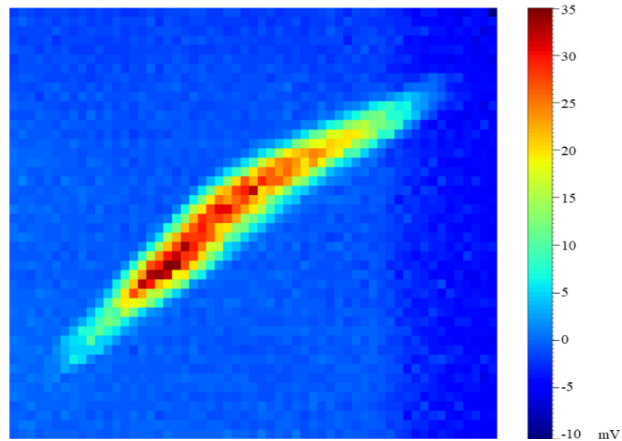

**Fig. 1** Two-dimensional transverse phase-space distribution measured with the IPHC emittance scanner. The pixelized grayscale image corresponds to a scan performed in one transverse plane, with a typical resolution of approximately 0.1 mm in position and 1 mrad in angle. The field of view is centered on the beam core and its surrounding halo. Background noise is distributed quasi-uniformly across the image and limits halo visibility, particularly in regions affected by optical aberrations (upper right and lower left corners)

## 2.2 Noise identification

Noise is an inherent and often detrimental byproduct of image-acquisition and image-processing systems. Its presence can significantly degrade not only the perceptual quality of an image but also the reliability of high-level computer-vision and quantitative-analysis tasks that depend on accurate pixel information. For this reason, noise suppression has long been a central challenge in the field of image denoising. In a broad sense, noise encompasses any perturbation external to the object under observation that introduces spurious or misleading pixel intensities, thereby reducing the fidelity of the acquired data. In practice, noise manifests as random or structured fluctuations in pixel values that obscure the true underlying signal and compromise the extraction of meaningful physical information. From the standpoint of signal theory, noise can be modeled as an additive or multiplicative component superimposed on the desired measurement, typically represented as a stochastic process with characteristic statistical properties. Its impact is particularly critical in scientific and industrial imaging applications, where the preservation of fine structural details, accurate intensity distributions, and reproducible measurements is essential for reliable diagnostics. The fundamental difficulty lies in distinguishing noise from genuine signal features, especially when the noise distribution is unknown, non-stationary, or varies dynamically with experimental conditions.

In the context of beam diagnostics for particle accelerators and other high-precision measurement environments, the consequences of noise extend far beyond visual degradation. Noise propagates into higher-level tasks such as contour detection, phase-space reconstruction, and emittance evaluation, thereby affecting the accuracy of key operational parameters. Robust denoising strategies are therefore indispensable to ensure that reconstructed images faithfully represent the underlying physical phenomena and support reliable machine tuning, loss mitigation, and operational safety. One of the persistent challenges in this domain is the absence of prior knowledge about the noise distribution. Experimental conditions often fluctuate, background contributions vary with time and location, and parasitic signals may arise from beam–instrument interactions. These factors hinder the development of universal denoising models and complicate the calibration of diagnostic devices. Classical statistical methods remain widely used across accelerator facilities to extract information from beam-diagnostic images; however, they exhibit significant limitations when confronted with noisy signals, irregular beam cross-sections, non-elliptical or non-Gaussian phase-space distributions, non-uniform density profiles, or the presence of contaminating secondary beams [21-27]. Recent studies have underscored the lack of flexibility of conventional filtering techniques, particularly under low signal-to-noise ratio (SNR) conditions. Traditional filters often struggle to suppress noise while preserving the edges and fine structures of regions of interest (ROI), leading to information loss and distortions in the reconstructed beam profile. Such shortcomings directly affect the accuracy of statistical emittance measurements (RMS), which are highly sensitive to the integrity of the underlying intensity distribution. Mischaracterization of beam properties can in turn compromise the matching between beam emittance and machine acceptance, ultimately becoming a source of malfunction or inefficiency in large-scale accelerator installations. The most common approach for RMS emittance determination relies on applying an arbitrary threshold to separate useful signal from background noise. However, even modest noise levels or the absence of minimal halo signals below $10^{-4}$ of the total beam intensity can introduce substantial bias, with RMS errors exceeding 100% in extreme cases. Conventional dark-field subtraction method, which records sensor noise in the absence of beam exposure, is insufficient because the beam's interaction with the scanner itself generates additional noise components that cannot be captured by dark frames alone. The inability to properly account for the beam halo, combined with limited measurement resolution characterized by high intensities in the beam core and rapidly decreasing intensities in the halo transition region remains a recurring challenge. Furthermore, background noise is not constant: it varies over time, across experimental campaigns, and depending on the instrument's location and installation environment. This variability complicates the definition of a single, reliable noise model and underscores the need for adaptive, data-driven denoising strategies [28-33]. The increasing complexity of measurement systems calls for more sophisticated approaches capable of extracting robust information from noisy, heterogeneous, and dynamically evolving data while preserving defects, anomalies, and details of the contour of the region of interest.

# 3 Methods

## 3.1 Data management and analysis

In this work, all datasets were acquired using the same high-resolution emittance scanner [34], with 2D grey-scaled images (≈ 500x500 pixels), and beams delivered from six low-energy accelerator facilities. The diversity of beam conditions provides a rich but challenging environment for image-based diagnostics. The combination of heterogeneous datasets, fluctuating noise characteristics, and the need to preserve fine halo structures motivates the development of unsupervised deep-learning tools capable of adapting to each image individually. A comprehensive analysis of two-dimensional phase-space distributions was performed on a large corpus of emittance images to ensure that the artificial-intelligence-based denoising tool is trained and tested on representative, diverse, and high-quality inputs. A preliminary data scrutiny is essential for identifying corrupted samples, class imbalances, and systematic noise patterns that could otherwise introduce bias into the learning process. The first stage consisted of assembling an extensive dataset encompassing both typical and atypical beam configurations, including irregular shapes, defective structures, and images exhibiting a wide range of noise levels. This diversity is crucial for developing a model capable of generalizing across the heterogeneous conditions encountered in operational accelerator environments. The initial dataset comprised approximately 2000 noisy images collected from six different low-energy accelerator installations [35]. These facilities differ in beam characteristics, experimental setups, emittance-scanner configurations, and measurement procedures, resulting in a broad variety of phase-space distributions. A substantial fraction of the initial batch, images lacking beam signal, affected by incorrect acquisition settings, exhibiting poor resolution, or containing severely off-center beams, were automatically rejected by a dedicated sorting algorithm. Ultimately, only about 10% of the initial dataset were retained as valid and informative samples. From this curated subset, representative images from each facility were selected to characterize their distinctive features. A series of standard data-analysis procedures was then applied, including basic statistical evaluations, assessments of noise and signal distributions, analyses of beam-intensity variations, and correlation studies. Frequency-domain analyses, such as Fast Fourier Transform (FFT), were also performed to identify periodic or structured noise components. As part of the preprocessing stage, the dataset was compared against several theoretical noise models including Gaussian, Uniform, Poisson, Speckle, Rician, Salt-and-Pepper, Exponential, and Gamma distributions to assess their compatibility with known statistical behaviors. Basic statistical descriptors were computed for both signal and noise regions, including pixel count, image dimensions (width × height), and intensity metrics such as minimum, maximum, mean, and standard deviation. Negative pixel values were retained when present, as they may arise from electronic noise, baseline drift, or subtraction artifacts commonly encountered in physical measurement systems. In a second processing step, signal and noise components were separated using a predefined intensity threshold, enabling their independent characterization. The resulting particle-position and angular distributions did not exactly conform to any standard theoretical model. However, increasing the threshold progressively yielded distributions that approached Gaussian behavior, consistent with the dominance of the beam core at higher intensity levels. No significant correlations were detected among the noise components, suggesting that the noise is largely unstructured and stochastic, see Fig. 2. Due to the substantial variability in image properties including size, beam shape, noise distribution, spatial scale, resolution, and region-of-interest centering, global normalization across the entire dataset proved infeasible. This heterogeneity reinforced the choice of an unsupervised learning strategy, in which denoising is performed on a per-image basis without relying on a universal normalization scheme. Such an approach is particularly well suited to accelerator diagnostics, where measurement conditions can vary widely and unpredictably across facilities and operational regimes.

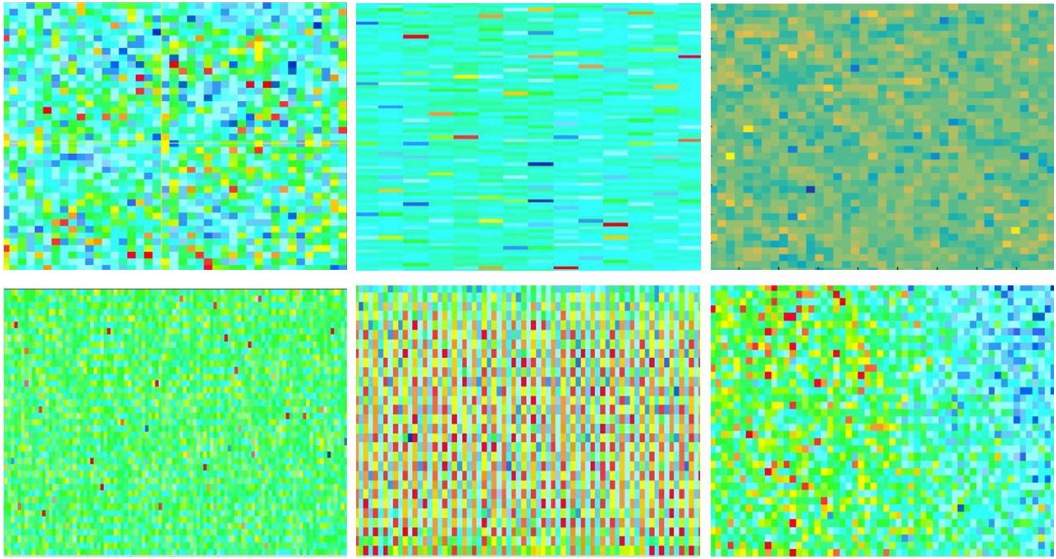

**Fig. 2** Noise identity portraits. Images of beam‑ scanner acquisitions from six different experimental setups obtained using the conventional dark‑ field subtraction procedure. Each image shows the two‑ dimensional transverse phase‑ space distribution of noise recorded in the absence of beam exposure (pixelized images of emittance figure expressed in mm·mrad unit, not represented). The colored display represents the equivalent beam‑ current intensity in the nanoampere range and is used solely to enhance contrast and readability

## 3.2 Image processing

Deep learning has transformed modern computer vision by dramatically reducing error rates in tasks such as object recognition, with exponential progress observed since 2010 [36]. While human interpretation benefits from experience, abstraction, and contextual reasoning, AI systems excel through their capacity to process vast quantities of data simultaneously and to extract patterns that may be imperceptible to human observers. This complementarity is particularly relevant for accelerator diagnostics, where image data are heterogeneous, noisy, and often lack clean reference labels. For beam‑ diagnostic applications, the diversity of infrastructures, operating environments, instrumentation, and beam conditions, combined with the absence of denoised or annotated ground‑ truth images makes supervised learning approaches impractical. The variability of noise sources, the lack of a reliable noise model, and the intrinsically low signal‑ to‑ noise ratio (SNR) of halo measurements further reinforce the need for flexible, unsupervised image‑ processing methods. In this context, the problem can be framed as a binary or ternary classification task involving three categories: foreground (signal), background, and the boundary region (including any outliers) defining the region of interest (ROI). The beam core corresponds to the ROI and exhibits the highest pixel intensities, whereas the tails and halo contain much lower densities. Background noise may overlap with the halo signal, especially in low‑ intensity experimental configurations, complicating the identification of the core–halo transition and limiting the achievable resolution. All images produced during our experiments are acquired in grayscale and processed at the pixel level using third‑ rank tensors, enabling the application of inverse‑ problem techniques such as classification, clustering, segmentation, pattern recognition, and denoising. Color maps are used solely for visualization purposes to enhance contrast and readability, without altering the underlying data. This representation is well suited to convolutional architectures, which can exploit spatial correlations and local structures to distinguish meaningful beam features from noise.

## 3.3 DCNN model foundations

The model developed to restore and denoise the images of the two-dimensional phase-space distributions is a deep convolutional neural network (DCNN) with a U-shaped encoder–decoder structure and skip connections [35] inspired by the Deep Image Prior (DIP) framework [37-40]. Convolutional hourglass networks distribute computational units commonly referred to as neurons not as isolated entities but as spatially organized elements within feature maps. In these architectures, each convolutional layer comprises a set of learnable filters, where each filter generates a feature map whose individual pixels function as neurons sharing identical parameters through weight sharing. This design preserves spatial locality and enables efficient representation learning across the image domain. The hourglass structure further enhances this capability by symmetrically combining a contracting path, which progressively reduces spatial resolution while increasing feature dimensionality, with an expanding path that restores resolution and integrates multi‑scale information via skip connections. Consequently, neurons are distributed across multiple spatial scales and semantic depths, allowing the network to jointly capture fine‑grained details and global contextual cues. This stands in contrast to fully connected deep neural networks, where neurons are independent units with unique weights and no inherent spatial structure. The hourglass architecture's multi‑resolution processing and spatially coherent neuron organization make it particularly effective for tasks such as denoising, semantic segmentation, and other vision problems requiring precise localization and contextual reasoning.

Although images are displayed using colors, the network processes only single-channel data (grayscale). Colors appear solely due to visualization colormaps applied when plotting scalar values. The down sampling path consists of two encoder blocks that extract spatial features while progressively reducing the image resolution, see Fig. 3. Each encoder block applies multiple convolutional and sequentially filtering. Convolutional filters with $3 \times 3$ kernels are stacked one after another with number of feature maps remaining constant, and each convolution is followed by Batch Normalization and LeakyReLU activation. Stacking several small kernels increases the effective receptive field while keeping the number of parameters moderate and introducing additional nonlinearities. Spatial resolution is reduced after each encoder block using a convolution with stride $s = 2$ and kernel size $3 \times 3$. No pooling layers are used. Down sampling is performed exclusively by stride convolutions, which is a standard design choice in modern CNNs and in the DIP framework. The bottleneck is the deepest part of the network. At this stage, the feature maps have a large receptive field, allowing the network to capture global image structure. The decoder reconstructs the image by progressively increasing spatial resolution while refining the extracted features. Up sampling is performed using bilinear interpolation with a scale factor of 2. No transposed convolutions are used, which helps avoid checkerboard artifacts. At each resolution level, skip connections link the encoder to the decoder. They allow fine spatial details to bypass the bottleneck and be reintroduced during reconstruction. Before merging, encoder features pass through a $1 \times 1$ convolution (1 to 2 filters for the first skip connection, and 32 to 2 filters for the second one, both being added to the 32 filters of the concatenation). A $1 \times 1$ convolution mixes information across layers at each pixel without affecting spatial resolution. Its role here is to adapt and compress encoder features before merging. Decoder and skip features are merged by channel-wise concatenation: 32 (decoder) + 2 (skip) = 34 filters. This operation preserves spatial information while enriching the representation with fine details from the encoder. Each decoder block applies two convolutional layers sequentially: a $3 \times 3$ convolution with the filter number reduced from 34 to 32 (by spatial fusion and filter reduction), and a $1 \times 1$ convolution keeping 32 filters (map refinement without spatial mixing). The $1\times1$ convolution inside the decoder should not be confused with the final output layer. Its purpose is to reorganize and reweight feature maps after concatenation, improving representation quality and optimization stability. The final layer is a $1 \times 1$ convolution reducing the number of filters from 32 to 1. This layer performs a pixel-wise linear projection from the learned feature representation to the final output value. Although a $1 \times 1$ convolution is also used inside the decoder blocks, the output $1 \times 1$ convolution has a different role, it produces the final scalar value at each pixel. No sigmoid activation is applied, so the output is linear. This allows the network to represent real-valued physical quantities (in mV).

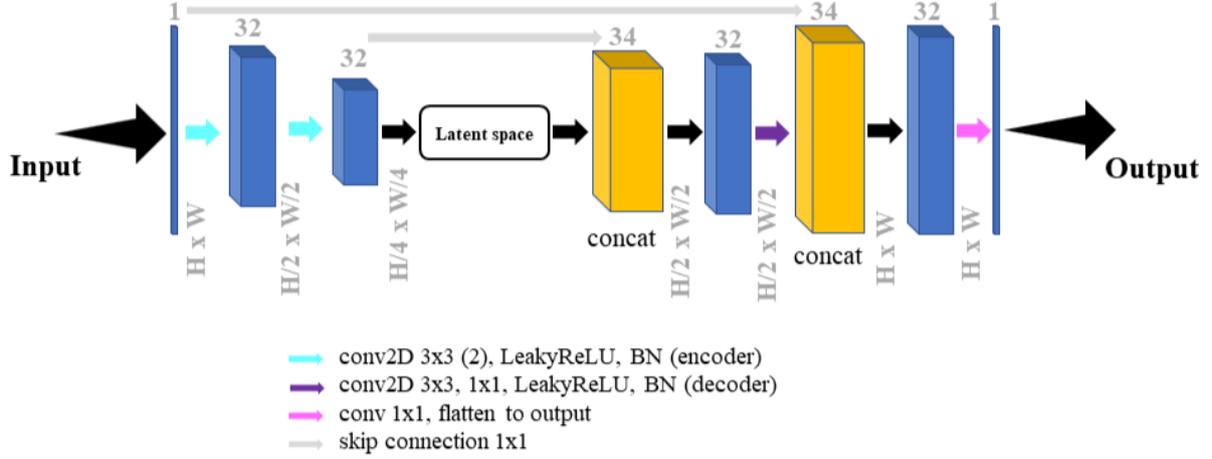

**Fig. 3** Architecture of the DCNN model (U‑NET structure) with two convolution blocks for encoding and down sampling, and two for decoding and up sampling, two layers of 32 filters each, kernels 3x3, strides 2, reflection padding operations, 2 skip connections, bilinear up convolutions (scale factor 2), LeakyReLU activation function, batch normalization (BN), reflection padding, no pooling, no sigmoid at output

## 4 Results

Ensuring that the model behaves consistently with the underlying physics requires evaluation metrics that retain direct physical interpretability, even though it is not possible to fully specify all desirable behaviors of the target distributions in advance. Two complementary metrics were therefore employed: the one‑ dimensional beam‑ intensity profile (DC distribution), see Fig. 4, and the two‑ dimensional RMS emittance derived from the transverse phase‑ space distribution. To quantify the latter, the RMS‑ ellipse surface is calculated at each iteration and used as an evaluation criterion. The evolution of the reconstructed beam area exhibits a characteristic pattern: a stable region following a local minimum corresponds to the optimal physical value, see Fig. 5. Beyond this point, the apparent beam area increases again, indicating the reinjection of noise into the reconstructed image and the emergence of false positives. The transition between these regimes is smooth and marked by a shallow slope. To support future systematic studies, an index was introduced to quantify the physical RMS emittance; its detailed formulation and assessment will be presented in a forthcoming publication. In the present work, a dedicated procedure was developed to align model inferences with physical expectations. Specifically, we analyzed the iteration count at which the model terminates under early stopping (ES) using a hybrid condition combining several scoring functions. This ES‑ based iteration number was then compared with the optimal iteration count obtained by tracking the evolution of the beam area during an extended run without ES, carried out up to 100 000 iterations. Close agreement between the two values indicates consistency between the ES‑ driven stopping criterion and the physically motivated emittance‑ based optimum. In cases where discrepancies were observed, model parameters and hyperparameters were adjusted, and the evaluation metrics refined, until a satisfactory balance between the two approaches was achieved.

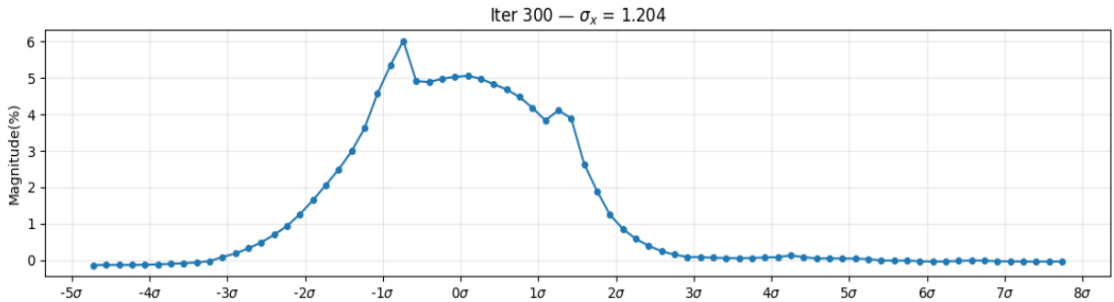

**Fig. 4** Beam‑ profile reconstruction with extended radial resolution. Reconstructed transverse beam profile obtained after data cleaning with the DCNN model shows measurable amplitudes extending beyond seven standard deviations. The reconstructed profile is used as a metric for the early‑ stopping procedure, which identifies the optimal iteration count by tracking the evolution of the beam‑ area criterion

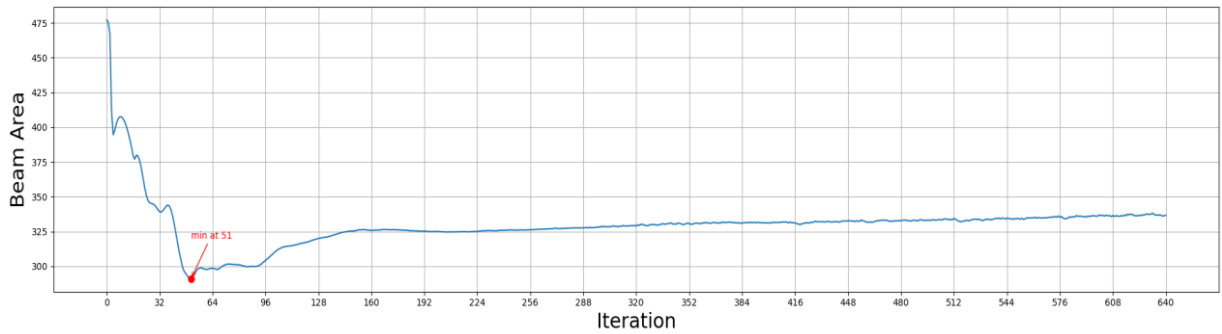

**Fig.5** Evolution of the beam area in two‑ dimensional phase space and arbitrary unit during the optimization process (640 iterations). The beam‑ area metric, derived from the RMS emittance and weighted by a dedicated index, is used to monitor and control model overfitting. The curve exhibits a local minimum followed by a gradual increase and a subsequent quasi‑ stable region, reflecting the transition from optimal reconstruction to noise reinjection in the DCNN output

After image reconstruction, the effective diagnostic resolution is significantly enhanced, see Fig. 6. Transverse amplitudes become detectable at radii beyond seven standard deviations from the beam core, and the system is able to resolve particle populations with local densities below $10^{-4}$ of the total beam intensity. This capability enables the faithful reconstruction of the outermost halo regions with a dynamic range that, to our knowledge, has not previously been achieved using this class of emittance scanner. Such extended sensitivity provides access to quantitative halo characterization essential for high‑ precision accelerator operation and for the development of next‑ generation beam‑ diagnostic tools.

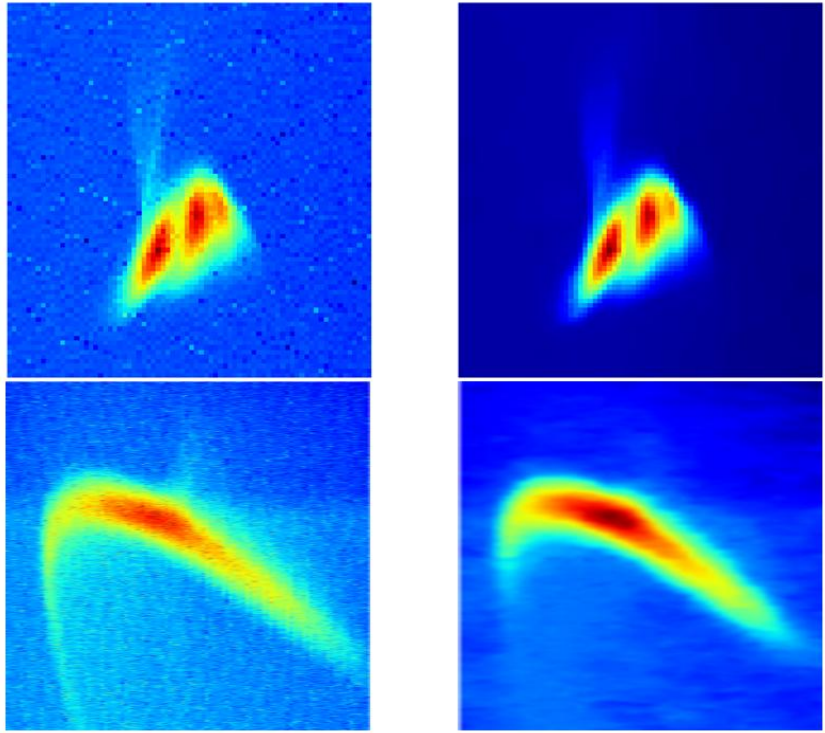

**Fig. 6** Restored and denoised emittance images. Noisy transverse phase‑ space scans are shown on the left, and the corresponding unsupervised DCNN‑ cleaned reconstructions on the right. The denoising process preserves the beam halo while effectively removing artifacts and background contamination, thereby enhancing the visibility of low‑ intensity structures. Results obtained after a few hundred training iterations correspond to a few minutes of processing on a CPU‑ based lap top

## 5 Discussion

The DCNN framework developed in this work, built upon a U‑Net architecture and complemented by new early‑stopping strategies, proves particularly effective for the reconstruction of both natural images and the complex distributions characteristic of particle‑accelerator beam dynamics. During training, the network progressively adjusts its weights to recover the underlying structure of a noisy image before it begins to fit the noise itself. This intrinsic property enables efficient denoising of emittance figures originating from a wide variety of beams and facilities, without requiring external datasets or paired clean/noisy images. The method preserves fine details and local textures with high fidelity, even in the absence of ground truth or explicit supervision. Early stopping plays a central role in preventing noise reinjection and the model exhibits a marked sensitivity to architectural choices and hyperparameter settings. To mitigate overfitting, self‑supervision strategies based on regularization and noise‑level constraints were introduced [35], significantly improving training stability, see Appendix. When the model is run for 10 000 iterations, the ES termination point shows a strong correlation with the evolution of the RMS emittance, revealing a clear relationship between the ES iteration count and the physical behavior of the reconstructed beam. This correlation indicates that the combination of DCNN reconstruction and ES constitutes a promising approach for enhancing beam diagnostics and improving the accuracy of emittance measurements. Nevertheless, scaling the framework to larger image datasets introduces new computational and methodological challenges, particularly regarding memory requirements, convergence behavior, and the management of heterogeneous noise conditions. The encouraging results obtained here motivate more systematic investigations and the establishment of a comprehensive benchmarking framework. A particularly relevant direction for future work is a controlled variational study in which beam parameters specifically diameter, intensity, and resolution are incrementally modified. Such perturbations would generate large sets of images spanning multiple configurations. Although inherently time‑consuming, this procedure would produce paired datasets designed to challenge the model with diverse representational conditions. These datasets could then serve as a foundation for training supervised models, enabling rigorous comparative evaluations and further improving the robustness of AI‑based beam‑diagnostic tools.

## 6 Conclusion

The study introduces an unsupervised deep‑convolutional neural‑network framework that markedly improves image denoising and restoration for beam‑diagnostic applications. By reconstructing emittance images including the halo extending from the beam core to the outermost transverse amplitudes, the method achieves a substantial enhancement in scan resolution through effective noise suppression and reliable extraction of weak signals embedded in the background. These results were obtained under particularly challenging conditions: very low signal‑to‑noise ratios, heterogeneous and irregular beam shapes, non‑Gaussian transverse distributions, and small, non‑annotated datasets. The achievable resolution is significantly extended, with measurable transverse amplitudes exceeding seven standard deviations radially in the most favorable cases. At this sensitivity level, the system resolves particle populations with local densities below $10^{-4}$ of the total beam intensity, enabling a quantitative representation of the halo with a precision that, to our knowledge, has not previously been reached using this type of emittance scanner. Notably, the framework enabled the identification of a halo structure that had not been observed at one of the pilot installations. The proposed DCNN approach relies on a standard U‑Net architecture augmented with carefully designed early‑stopping strategies to prevent overfitting, a set of adapted evaluation metrics, and a combined loss function that integrates quantitative physical criteria with qualitative perceptual considerations derived from extensive heuristic analysis.

    A key feature of the method is its frugality: it requires minimal computational resources, operates entirely on CPUs, and does not depend on cloud infrastructures, data centers, or internet connectivity. This lightweight design makes the framework particularly suitable for accelerator environments where ground‑truth data and large training sets are unavailable, while also contributing to a reduced carbon footprint consistent with sustainable‑development objectives. The results demonstrate that unsupervised deep‑learning methods can serve as powerful enablers for next‑generation beam diagnostics, offering enhanced resolution and interpretability under operationally realistic conditions. In addition, the image‑based approach allows an agnostic solution, meaning it does not depend on beam energy, scanner class, or nature of the measurement distribution. The approach opens the way for broader deployment of AI‑based tools in accelerator facilities and motivates further work toward systematic benchmarking, controlled dataset generation, and the integration of physics‑informed learning strategies.

## Acknowledgement

The authors did not receive support from any organization for the submitted work.

# **Appendix**  Image of emittance figure restoration

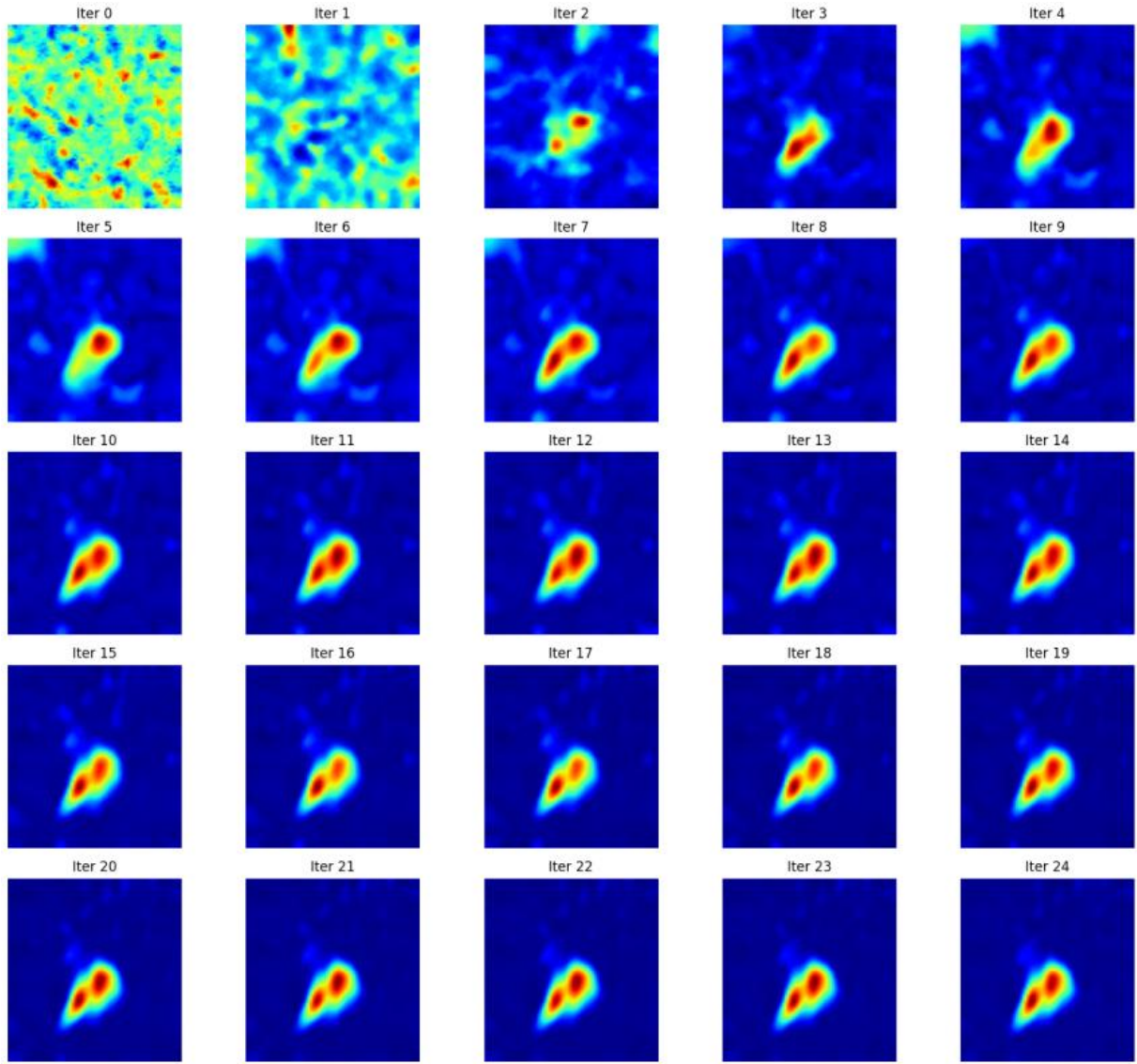

Progressive restoration of emittance images over 24 iterations. Sequence of reconstructed transverse phase‑ space images obtained from an initial Gaussian‑ noise input taken as a regularization constraint (upper left) and a corrupted scan (shown in Fig. 6). Across iterations, the DCNN progressively suppresses noise and replaces it with semantically meaningful structures, yielding a cleaner and physically consistent emittance representation